\documentclass[a4paper]{jpconf}
\usepackage{graphicx,epsfig,amsmath,amssymb,color}

\newcommand{\mhh}{m_{hh}}

\begin{document}
\title{SMEFT truncation effects in Higgs boson pair production at NLO QCD}

\author{Gudrun Heinrich, Jannis Lang}

\address{Institute for Theoretical Physics, Karlsruhe Institute of Technology, 76131 Karlsruhe, Germany}

\ead{gudrun.heinrich@kit.edu, jannis.lang@kit.edu}

\begin{abstract}
We present results for Higgs boson pair production in gluon fusion at
next-to-leading order in QCD, including effects of anomalous couplings within  Standard Model
Effective Field Theory (SMEFT). In particular, we investigate truncation effects of the SMEFT series, 
comparing different ways to treat powers of dimension-six operators and double operator insertions.
\end{abstract}

\section{Introduction}

Higgs boson pair production in gluon fusion offers the possibility to measure the trilinear Higgs boson
self-coupling and therefore to verify whether the form of the Higgs
potential assumed in the Standard Model (SM)
is correct.
Deviations from this form, manifesting themselves in anomalous Higgs boson
self-couplings,  would be a clear sign of new physics and most likely
would come along with other non-SM Higgs couplings.
Therefore it is important to control the uncertainties of the theory
predictions in simulations that include anomalous couplings.
The theoretical uncertainties have various sources, the dominant ones in the SM being uncertainties related to the top quark mass renormalisation scheme.
Theory predictions with full top quark mass dependence are available
at NLO
QCD~\cite{Borowka:2016ehy,Borowka:2016ypz,Baglio:2018lrj,Baglio:2020ini}
and have been included in calculations where higher orders have been
performed in the heavy top
limit~\cite{Grazzini:2018bsd,Chen:2019lzz,Chen:2019fhs}, thus reducing
the scale uncertainties and the uncertainties due to missing top quark
mass effects, such that the top mass scheme uncertainties currently
constitute the main uncertainties~\cite{Baglio:2020wgt} of the SM predictions.


Going beyond the SM description of the process $gg\to HH$, considering
in particular effective field theory (EFT) parametrisations of new
physics effects, new uncertainties arise, coming mainly from the truncation
of the EFT expansion.

In the following we will present results at NLO SMEFT for this process, including also double operator insertions.
Our implementation allows us to investigate various scenarios of truncation and to assess the related uncertainties.
For more details we refer to Ref.~\cite{Heinrich:2022idm}.

\section{Effective field theory descriptions of Higgs boson pair production}

\subsection{HEFT and SMEFT}

In Standard Model Effective Field Theory
(SMEFT)~\cite{Grzadkowski:2010es,Brivio:2017vri}, an effective
description of unknown interactions at a new physics scale $\Lambda$
is constructed as an expansion in inverse powers of $\Lambda$, with
operators $\mathcal{O}_i$ of canonical dimension larger than four and
corresponding Wilson coefficients $C_i$,
\begin{equation}
\mathcal{L}_\text{SMEFT} = \mathcal{L}_\text{SM} + \sum_{i}\frac{C_i^{(6)}}{\Lambda^2}\mathcal{O}_i^{\rm{dim6}} +{\cal O}(\frac{1}{\Lambda^3})\; .
\label{eq:Ldim6}
\end{equation}
In SMEFT it is assumed that the physical Higgs boson is part of a
doublet transforming linearly under $SU(2)_L\times U(1)$.

Higgs Effective Field Theory (HEFT) instead is based on an expansion
in terms of loop orders, which also can be formulated in terms of
chiral dimension counting~\cite{Buchalla:2013eza,Krause:2016uhw}. The expansion parameter is
given by $f^2/\Lambda^2\simeq \frac{1}{16\pi^2}$, where $f$ is a
typical energy scale at which the EFT expansion is valid (for example
the pion decay constant in chiral perturbation theory),
\begin{align}
  {\cal L}_{d_\chi}={\cal L}_{(d_\chi=2)}+\sum_{L=1}^\infty\sum_i\left(\frac{1}{16\pi^2}\right)^L c_i^{(L)} O^{(L)}_i\;.
  \label{eq:loop_expansion}
  \end{align}

The SMEFT Lagrangian is typically given in the so-called Warsaw
basis~\cite{Grzadkowski:2010es}, the terms relevant to the process
$gg\to HH$ read
\begin{equation}
\begin{split}
\Delta\mathcal{L}_{\text{Warsaw}}&=\frac{C_{H,\Box}}{\Lambda^2} (\phi^{\dagger} \phi)\Box (\phi^{\dagger } \phi)+ \frac{C_{H D}}{\Lambda^2}(\phi^{\dagger} D_{\mu}\phi)^*(\phi^{\dagger}D^{\mu}\phi)+ \frac{C_H}{\Lambda^2} (\phi^{\dagger}\phi)^3 \\ &+\left( \frac{C_{uH}}{\Lambda^2} \phi^{\dagger}{\phi}\bar{q}_L\phi^c t_R + h.c.\right)+\frac{C_{H G}}{\Lambda^2} \phi^{\dagger} \phi G_{\mu\nu}^a G^{\mu\nu,a}\;.  \label{eq:warsaw}
\end{split}
\end{equation}
The dipole operator $\bar{{\cal O}}_{uG}$ is not included here because it
can be shown that it carries an extra loop suppression factor $1/16\pi^2$ relative to the other contributions if weak coupling to the heavy sector is assumed~\cite{Buchalla:2018yce,Arzt:1994gp}. In the strong coupling case an expansion in the canonical dimension only would not be the appropriate description.

The HEFT Lagrangian relevant to Higgs boson pair production in gluon
fusion can be parametrised by five a priori independent anomalous
couplings as follows~\cite{Buchalla:2018yce}
\begin{align}
\Delta{\cal L}_{\text{HEFT}}=
-m_t\left(c_t\frac{h}{v}+c_{tt}\frac{h^2}{v^2}\right)\,\bar{t}\,t -
c_{hhh} \frac{m_h^2}{2v} h^3+\frac{\alpha_s}{8\pi} \left( c_{ggh} \frac{h}{v}+
c_{gghh}\frac{h^2}{v^2}  \right)\, G^a_{\mu \nu} G^{a,\mu \nu}\;.
\label{eq:ewchl}
\end{align}

Expanding the Higgs doublet in eq. \eqref{eq:warsaw} around its vacuum
expectation value and applying a field redefinition for the physical
Higgs boson
\begin{align}
h\to h+v^2\frac{C_{H,kin}}{\Lambda^2}\left(h+h^2+\frac{h^3}{3}\right)\;,\label{eq:field_redefinition}
\end{align} 
with $\frac{C_{H,kin}}{\Lambda^2}:=\frac{C_{H,\Box}}{\Lambda^2}-\frac{1}{4}\frac{C_{HD}}{\Lambda^2}$, the Higgs kinetic term acquires its canonical form (up to ${\cal O}\left(\Lambda^{-4}\right)$ terms).
After that, relating the couplings through a comparison of the coefficients of the
corresponding terms in the Lagrangian leads to the expressions given
in Table~\ref{tab:translation}.
\begin{table}[htb]
\begin{center}
\begin{tabular}{ |c |c| }
\hline
HEFT&Warsaw\\
\hline
$c_{hhh}$ & $1-2\frac{v^2}{\Lambda^2}\frac{v^2}{m_h^2}\,C_H+3\frac{v^2}{\Lambda^2}\,C_{H,kin}$ \\
\hline
$c_t$ &  $1+\frac{v^2}{\Lambda^2}\,C_{H,kin} - \frac{v^2}{\Lambda^2} \frac{v}{\sqrt{2} m_t}\,C_{uH}$\\
\hline
$ c_{tt} $ & $-\frac{v^2}{\Lambda^2} \frac{3 v}{2\sqrt{2} m_t}\,C_{uH} + \frac{v^2}{\Lambda^2}\,C_{H,kin}$\\
\hline
$c_{ggh}$ &  $\frac{v^2}{\Lambda^2} \frac{8\pi }{\alpha_s} \,C_{HG}$ \\
\hline
$c_{gghh}$ &  $\frac{v^2}{\Lambda^2}\frac{4\pi}{\alpha_s} \,C_{HG}$ \\
\hline
\end{tabular}
\end{center}
\caption{Leading order translation between different operator basis choices.\label{tab:translation}}
\end{table}

However, it should be emphasized that a translation between the coefficients at Lagrangian level must be
applied with care.
The EFT parametrisations have a validity range limited by unitarity
constraints and the assumption that $C_i/\Lambda^2$ in SMEFT is a small
quantity. Furthermore there are relations between the coefficients in
SMEFT which are not present in HEFT.
Therefore a naive translation from HEFT (which is more general) to
SMEFT can lead out of the validity range for a given point in
the coupling parameter space, even if it is a perfectly valid
point in HEFT.

\subsection{SMEFT truncation}

Another delicate point in the EFT expansion is the question how to
treat terms with inverse powers of $\Lambda$ higher than two at cross
section level, i.e.  when squaring the amplitude.
These are terms related to squared dim-6 operators,  double
operator insertions in a single diagram and combinations thereof.
Related issues have been discussed recently in Ref.~\cite{Brivio:2022pyi}.

We now present a Monte Carlo program which allows us to study the truncation effects systematically.
In order to construct the different truncation options we first deconstruct the amplitude into three parts:
the pure SM contribution ($\text{SM}$), single dim-6 operator insertions ($\rm{dim6}$) and double dim-6 operator insertions ($\rm{dim6}^2$):
\begin{align}
{\cal M}=&\ 
\begin{aligned}\vcenter{\hbox{\includegraphics[page=8,scale=0.7]{figures/HH_presentation}}}\\[-6pt]\scalebox{0.1}{\textcolor{white}{dummy}}\end{aligned}
+\vcenter{\hbox{\includegraphics[page=9,scale=0.7]{figures/HH_presentation}}}
+\vcenter{\hbox{\includegraphics[page=10,scale=0.7]{figures/HH_presentation}}}
\nonumber\\
&\ +\vcenter{\hbox{\includegraphics[page=11,scale=0.7]{figures/HH_presentation}}}
+\vcenter{\hbox{\includegraphics[page=12,scale=0.7]{figures/HH_presentation}}}
+\dots
\nonumber\\
=&\ 
{\cal M}_\text{SM} + {\cal M}_{\rm{dim6}} + {\cal M}_{\rm{dim6}^2}\;,\label{eq:amplitude_expansion}
\end{align}
where $c^\prime$ denotes the corresponding coupling combination listed in Table~\ref{tab:translation}.
For the squared amplitude forming the cross section, we consider four possibilities to choose which parts of $|{\cal M}|^2$ from eq. \eqref{eq:amplitude_expansion} may enter:
\begin{align}
\sigma \simeq \left\{\begin{aligned}
&\ \sigma_\text{SM} + \sigma_{\text{SM}\times \rm{dim6}}
\\
&\ \sigma_{\left(\text{SM}+\rm{dim6}\right)\times \left(\text{SM}+\rm{dim6}\right)}  
\\
&\ \sigma_{\left(\text{SM}+\rm{dim6}\right)\times \left(\text{SM}+\rm{dim6}\right)}  + \sigma_{\text{SM}\times \rm{dim6}^2} 
\\
&\ \sigma_{\left(\text{SM}+\rm{dim6}+\rm{dim6}^2\right)\times \left(\text{SM}+\rm{dim6}+\rm{dim6}^2\right)}\;.
\end{aligned}\right.\label{eq:truncation}
\end{align}
The first line is the first order of an expansion of $\sigma\sim |{\cal M}|^2$ in $\Lambda^{-2}$, the second term is the first order of an expansion of ${\cal M}$ in $\Lambda^{-2}$. 
The third line includes all terms of ${\cal O}\left(\Lambda^{-4}\right)$ coming from single and double dim-6 operator insertions, however it lacks the contribution at the same order from dim-8 operators and ${\cal O}\left(\Lambda^{-4}\right)$ terms following the field redefinition of eq. \eqref{eq:field_redefinition}.
The fourth line is the naive translation from HEFT to SMEFT using Table~\ref{tab:translation}.
Typically, only the first two options are used for predictions and measurements using SMEFT, since both are unambiguous wrt. basis change and gauge invariance, however there is still a debate about the recommendations for their application to experimental analyses~\cite{Brivio:2022pyi}.
Thus, we include all of the presented options in our calculation, which can serve to contrast different outcomes of the predictions.

\section{NLO Implementation into the event generator program {\tt POWHEG} and results}

\subsection{Parametrisation of the $gg\to HH$ total cross section}

Our implementation is based on the publicly available NLO HEFT code presented in Refs.~\cite{Heinrich:2019bkc,Heinrich:2020ckp},
converted to the SMEFT framework and extended such that the different options described in the previous section can be calculated, including NLO QCD corrections.


For the real emission, a modified {\tt GoSam}~\cite{Cullen:2014yla} version is built that splits the amplitude evaluation according to eq. \eqref{eq:amplitude_expansion} and is able to generate the squared amplitude with the truncation option which can be set by the user via an input variable.
For the generation of the {\tt GoSam} files in {\tt POWHEG}~\cite{Alioli:2010xd}, a model in UFO format~\cite{Degrande:2011ua} has been produced which specifies the anomalous couplings such that {\tt GoSam} is able to calculate the different contributions according to the chosen truncation option. The existing interface to {\tt POWHEG} has been modified to hand over event parameters to {\tt GoSam} such that the factor $\alpha_s$ between Higgs-gluon couplings in HEFT and SMEFT is evaluated at the correct energy scale.

The virtual part is based on grids encoding the virtual 2-loop amplitudes.
These grids can be used to reconstruct the amplitude for any given combination of anomalous couplings.
The $a_i$ listed below are defined as the coefficients in the representation of the sqared amplitude as a linear combination of all coupling combinations possible in HEFT at NLO QCD.
\begin{align*}
|{\cal M}_{BSM}|^2 =& a_1\cdot c_t^4+a_2\cdot c_{tt}^2+a_3\cdot c_t^2 c_{hhh}^2 + a_4\cdot c_{ggh}^2 c_{hhh}^2 + a_5\cdot c_{gghh}^2 + a_6\cdot c_{tt}c_t^2 + a_7\cdot c_t^3 c_{hhh}  \nonumber\\[-2pt]
+&a_8\cdot c_{tt} c_t c_{hhh} + a_9\cdot c_{tt} c_{ggh} c_{hhh} + a_{10}\cdot c_{tt} c_{gghh} +a_{11}\cdot c_t^2 c_{ggh} c_{hhh} + a_{12}\cdot c_t^2 c_{gghh}  \nonumber\\[-2pt]
+&a_{13}\cdot c_{t} c_{hhh}^2 c_{ggh} + a_{14}\cdot c_{t} c_{hhh} c_{gghh} + a_{15}\cdot c_{ggh} c_{hhh}c_{gghh} +a_{16}\cdot c_t^3 c_{ggh}  \nonumber\\[-2pt]
+&a_{17}\cdot c_t c_{tt}c_{ggh} + a_{18}\cdot c_{t} c_{ggh}^2 c_{hhh} + a_{19}\cdot c_{t} c_{ggh} c_{gghh} + a_{20}\cdot c_{t}^2 c_{ggh}^2  \nonumber\\[-2pt]
+&a_{21}\cdot c_{tt} c_{ggh}^2 + a_{22}\cdot c_{ggh}^3 c_{hhh} + a_{23}\cdot c_{ggh}^2 c_{gghh}\;.\label{eq:squared_amp}
\end{align*}
Since the first and the third truncation options of eq.~\eqref{eq:truncation} are expansions at cross section level,
and the fourth option is the direct translation from HEFT to SMEFT, for those cases the application of the translation of Table~\ref{tab:translation}, including all terms at the desired order in inverse $\Lambda$, is sufficient.
In the case of the second truncation option, some of the grids can be reused as well, but the determination of the coefficients needs more care, as there are additional combinations.
Note that we do not include RGE running of the couplings as we only consider NLO QCD corrections to the amplitude, which factorise.

\begin{figure}[h]
\begin{center}
\includegraphics[width=14pc,page=1]{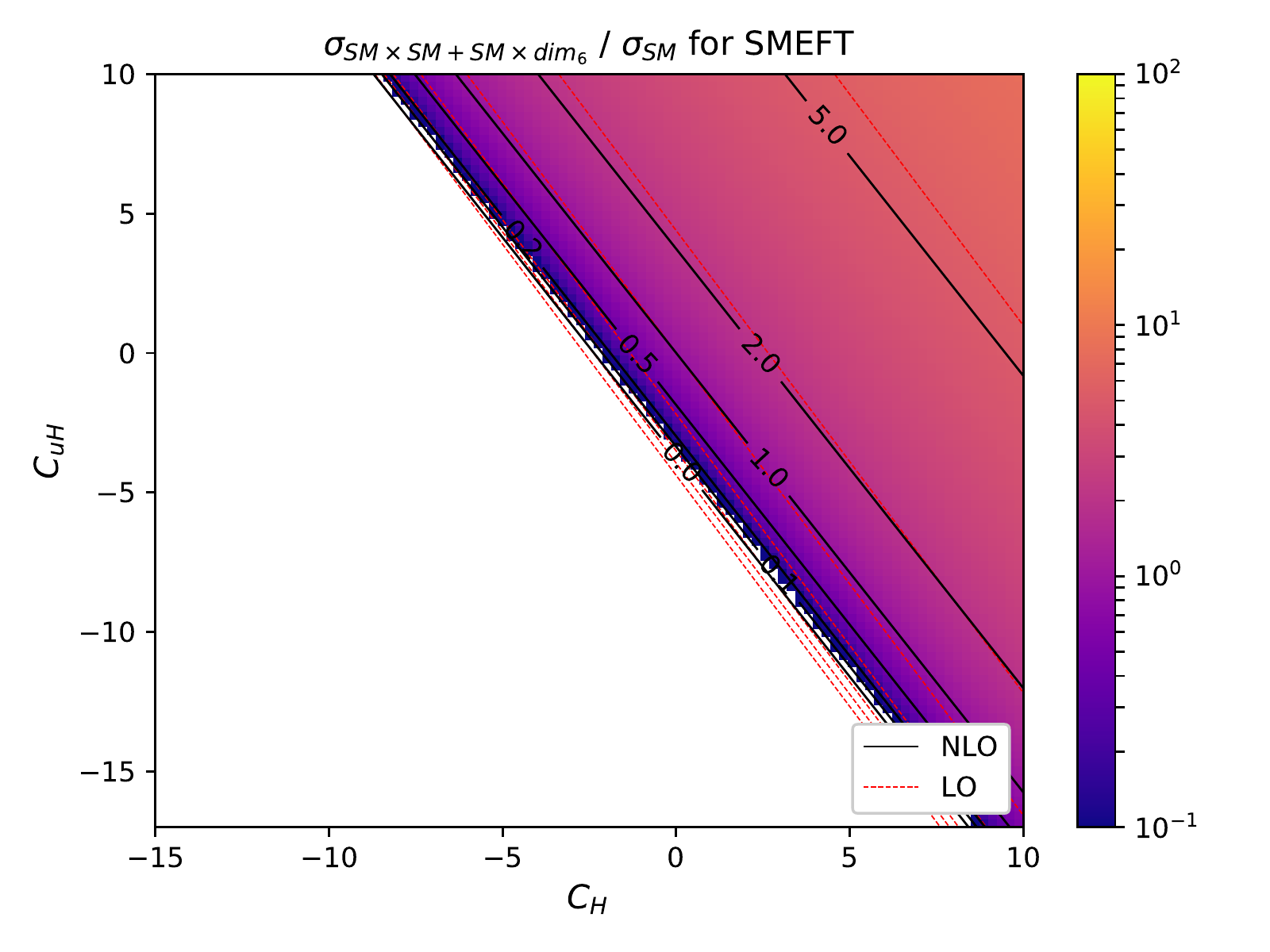}\hspace{2pc}%
\includegraphics[width=14pc,page=1]{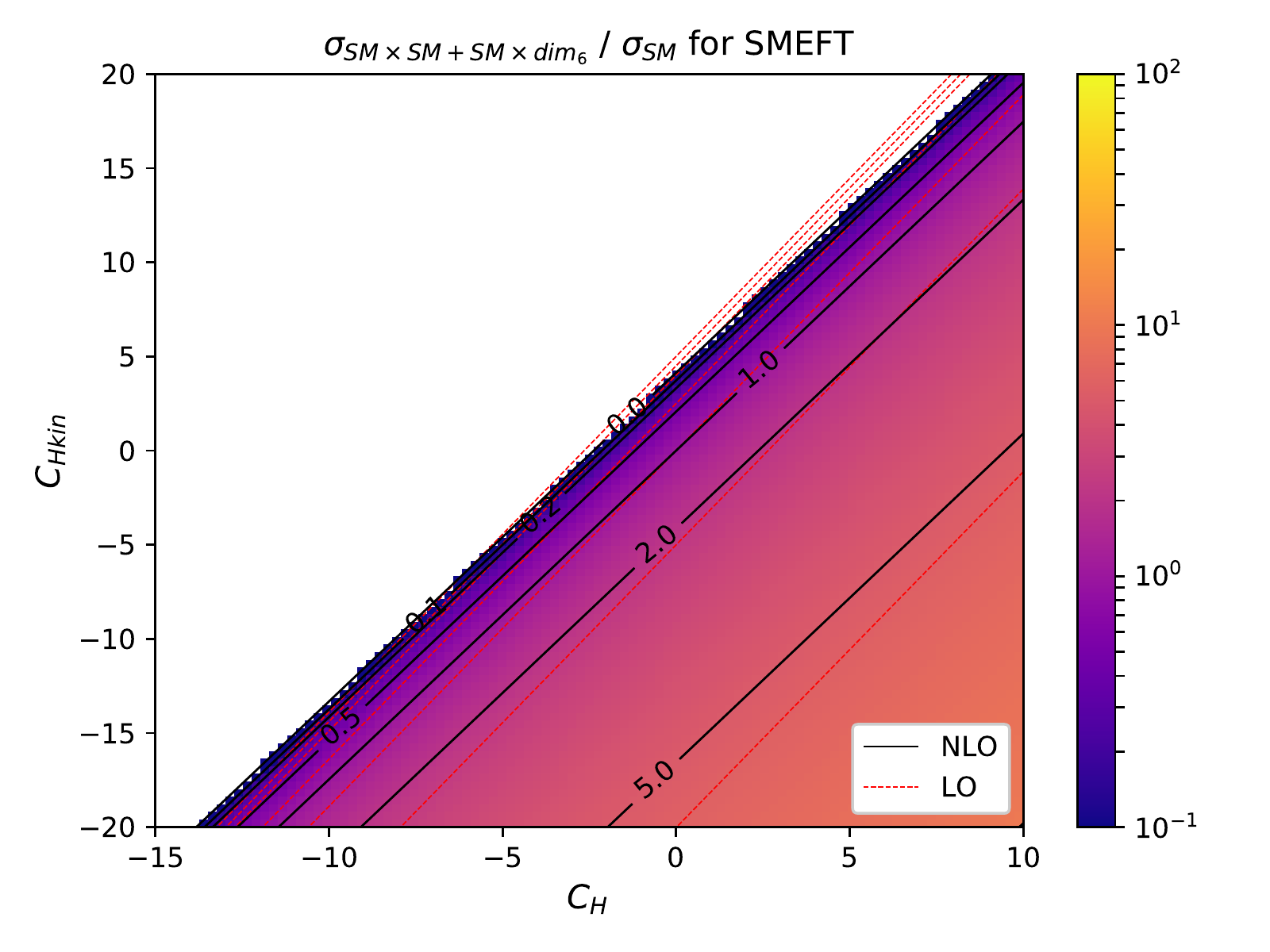}\hspace{2pc}%
\\
\includegraphics[width=14pc,page=2]{figures/plot_CH_CuH.pdf}\hspace{2pc}%
\includegraphics[width=14pc,page=2]{figures/plot_CH_CHkin.pdf}\hspace{2pc}%
    \caption{\label{heatmaps} Heat maps showing
    the dependence of the cross section on the couplings $C_H$, $C_{uH}$ (left) and
    $C_H$, $C_{H,kin}$ (right) with $\Lambda=1$\,TeV for different truncation options. Top: option 1 (linear dim-6), bottom: option 2 (quadratic dim-6).
  The white areas denote regions in parameter space where the corresponding cross section would be negative.}
\end{center}
\end{figure}


In Fig.~\ref{heatmaps} we show that the results for the total cross sections (normalised to the SM case) are substantially different between option 1 (linear dim-6, top) and option 2 (quadratic dim-6, bottom).
The white areas come from the fact that taking into account only linear dim6-contributions leads to negative cross sections over large parts of the parameter space. Furthermore, in the linear dim-6 case, there appears to be a completely flat direction in the observed parameter range for a combined variation of the respective Wilson coefficients in the diagrams. Flat directions are apparent in option 2 as well, however they correspond to an elliptic shape of equipotential lines due to the quadratic terms in the cross section.


\subsection{Investigation of truncation effects for the Higgs boson
  pair invariant mass distribution}

Now we turn to differential results, showing the effects of the different truncation options on the Higgs boson pair invariant mass distribution $\mhh$. We present results at two benchmark points, given in Table~\ref{tab:benchmarks},  which were derived analogously to \cite{Capozi:2019xsi} based on an analysis of characteristic shapes of the $\mhh$ distribution, but with the inclusion of current experimental constraints.
The upper panels show results for 
$\Lambda=1$\,TeV, the lower panels show results for the same point for  $\Lambda=2$\,TeV, for the different truncation options.
One can clearly see that
(a) the negative differential cross section values in the linear dim-6 case (blue) indicate that parameter points in anomalous coupling parameter space which are valid in HEFT can lead, upon naive translation, to parameter points for which the SMEFT expansion is not valid,
(b) destructive interference between different parts of the amplitude (e.g. box- and triangle-type diagams) can be enhanced or diminished depending on the truncation option,
(c) increasing $\Lambda$ reduces the differences between the results as they are smaller deformations of the SM parameter space.
In addition, we observe that the contribution from the interference of double dim-6 operator insertions with the SM appears to be subdominant in 
the example of benchmark point $1^\ast$,
 as can be seen by comparing truncation option 2 (orange) with option 3 (red), the latter including the double operator insertions.
We also should point out that the difference between HEFT (cyan) and SMEFT with truncation option 4 (green) is due to the scale dependence of $\alpha_s$, coming from the definition of $C_{HG}$ in the Warsaw basis,  see Table~\ref{tab:translation}.

\begin{table}[htb]
  \begin{center}
    \begin{footnotesize}
\begin{tabular}{ |c|c|c|c|c|c||c|c|c|c|c| }
\hline
\begin{tabular}{c}
benchmark \\
($^\ast=$ modified)
\end{tabular} & $c_{hhh}$ & $c_t$ & $ c_{tt} $ & $c_{ggh}$ & $c_{gghh}$ & $C_{H,\textrm{kin}}$ & $C_{H}$ & $C_{uH}$ & $C_{HG}$ & $\Lambda$\\
\hline
SM & $1$ & $1$ & $0$ & $0$ & $0$ & $0$ & $0$ & $0$ & $0$ & $1\;$TeV\\
\hline
$1^\ast$ & $5.105$ & $1.1$ & $0$ & $0$ & $0$ & $4.95$ & $-6.81$ & $3.28$ & $0$ & $1\;$TeV\\
\hline
$6^\ast$ & $-0.684$ & $0.9$ & $ -\frac{1}{6} $ & $0.5$ & $0.25$ & $0.561$ & $3.80$ & $2.20$ & $0.0387$ & $1\;$TeV\\
\hline
\end{tabular}
\end{footnotesize}
\end{center}
\caption{Benchmark points used for the invariant mass distributions. The benchmark points were derived analogously to \cite{Capozi:2019xsi}, but are slightly modified compared to the ones given in~\cite{Capozi:2019xsi}, to take into account current experimental constraints. The value of $C_{HG}$ is determined using $\alpha_s(m_Z)=0.118$. \label{tab:benchmarks}}
\end{table}

\begin{figure}[h]
\includegraphics[width=18pc,page=1]{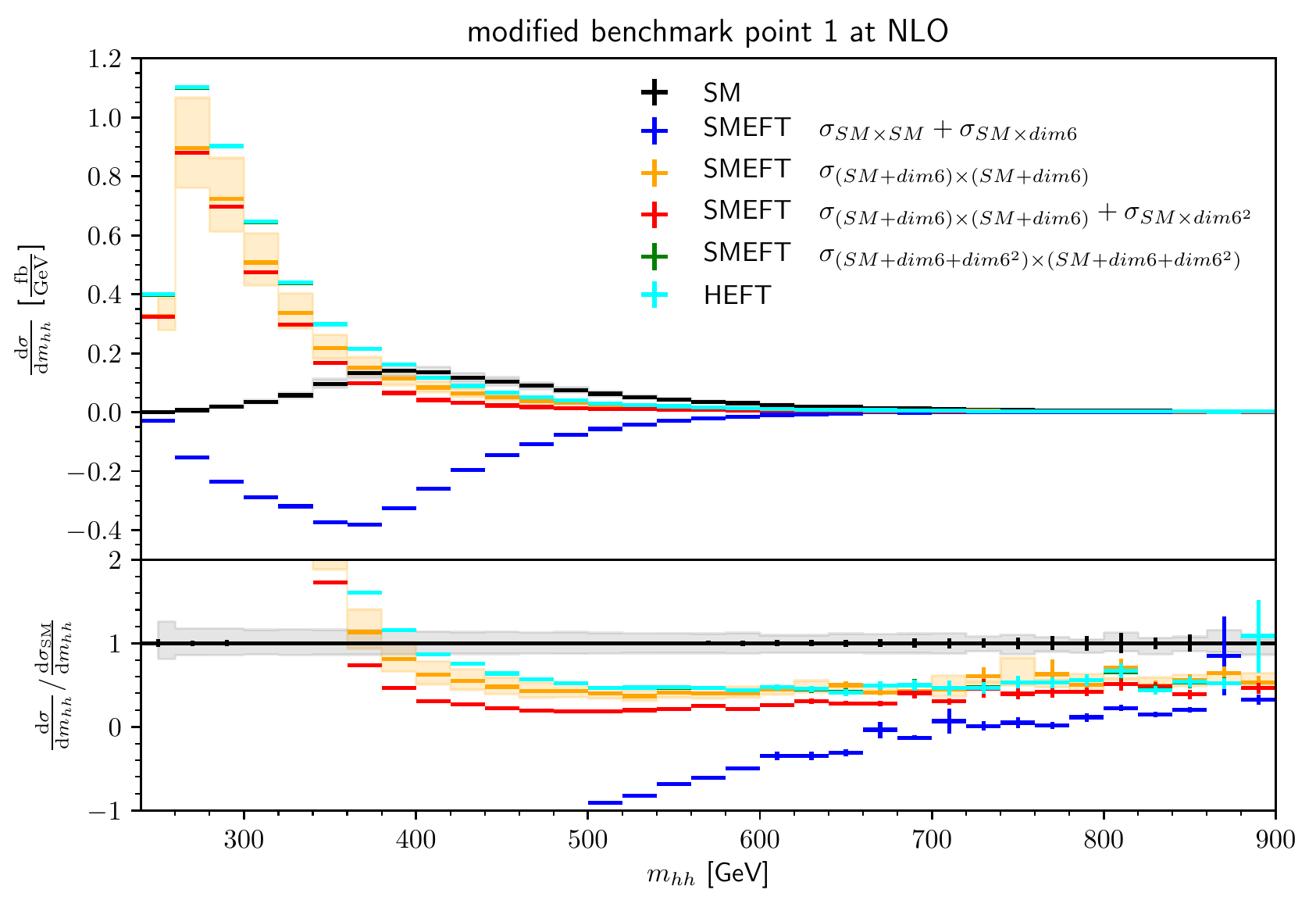}\hspace{2pc}%
\includegraphics[width=18pc,page=1]{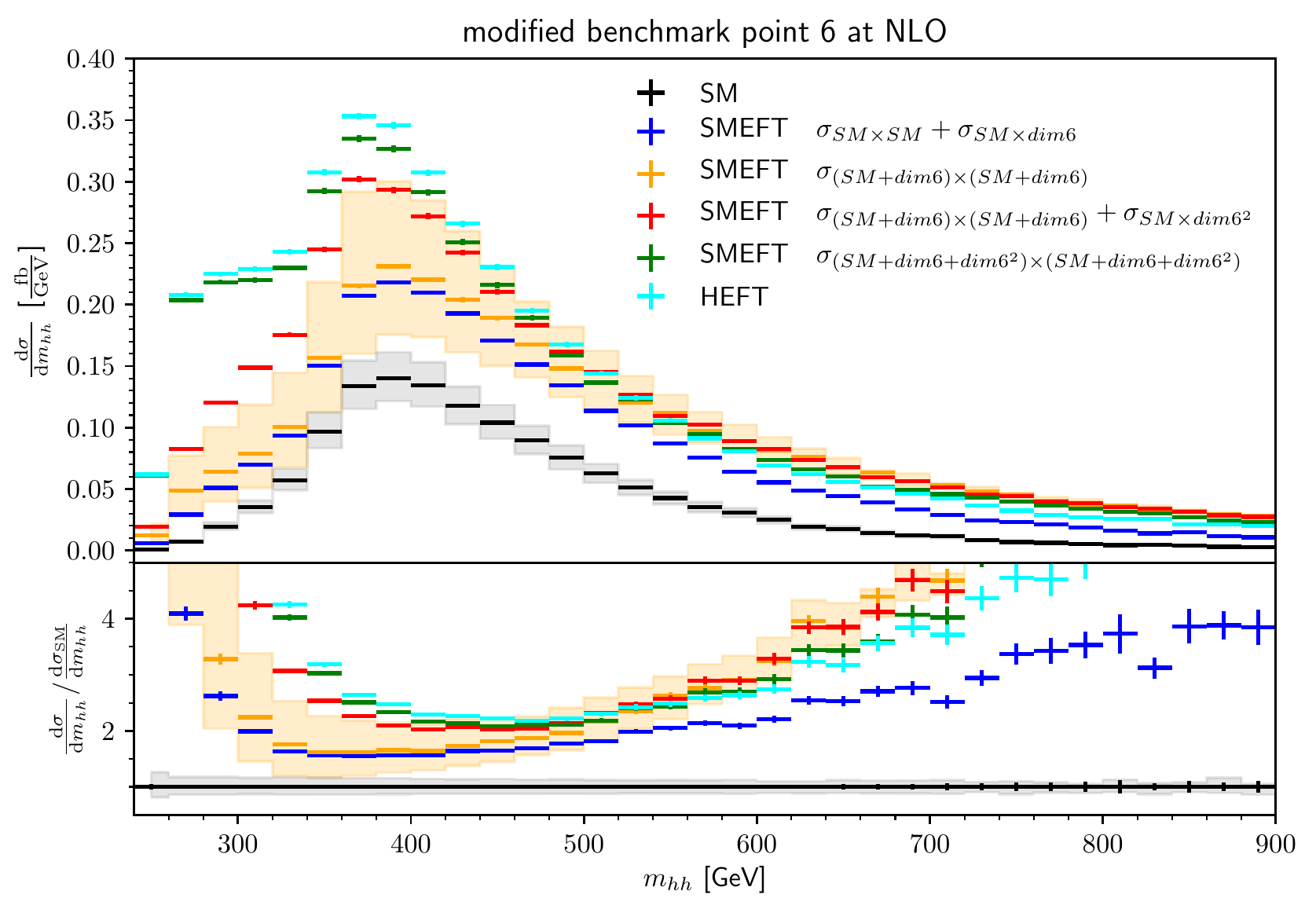}\hspace{2pc}%
\\
\includegraphics[width=18pc,page=1]{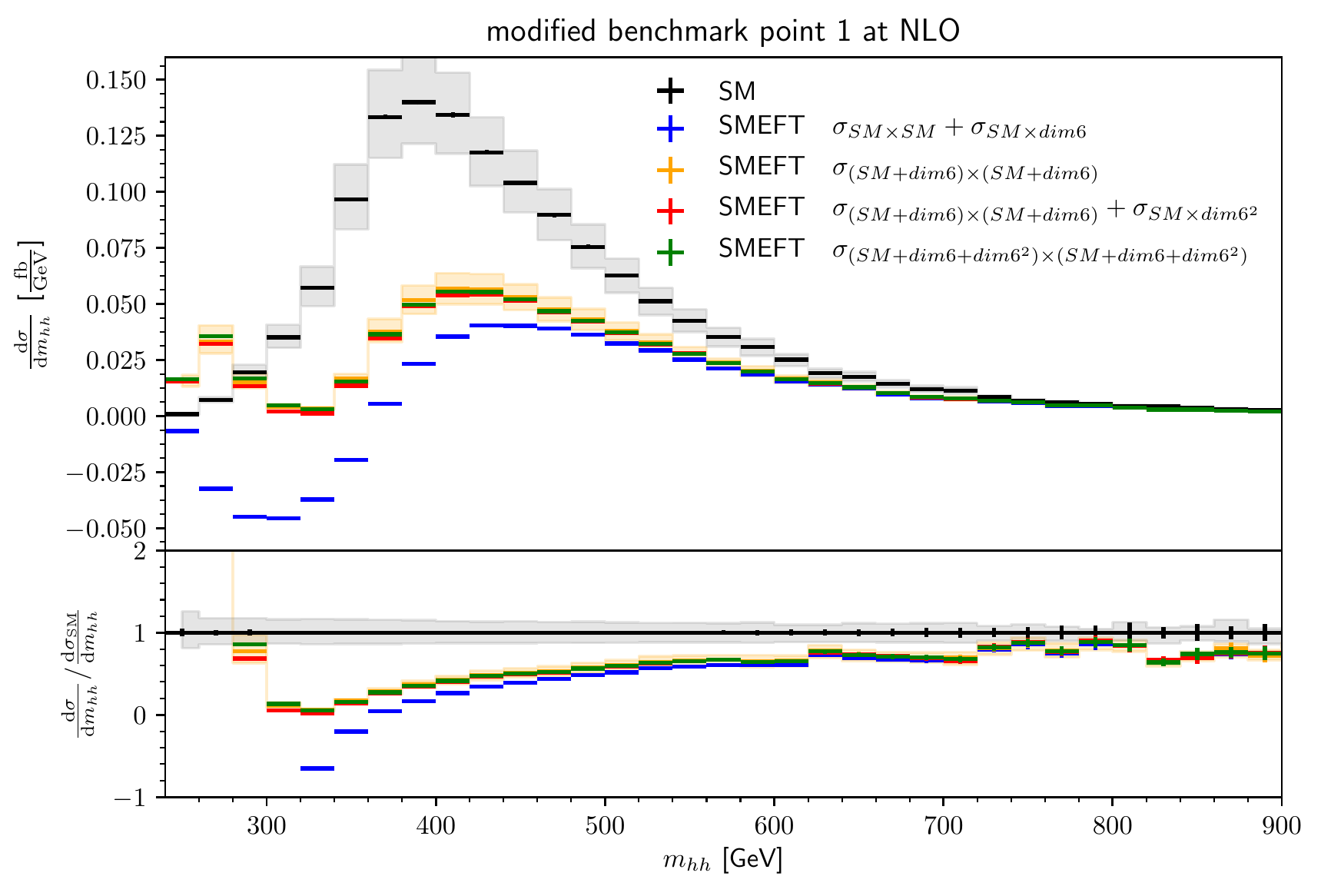}\hspace{2pc}%
\includegraphics[width=18pc,page=1]{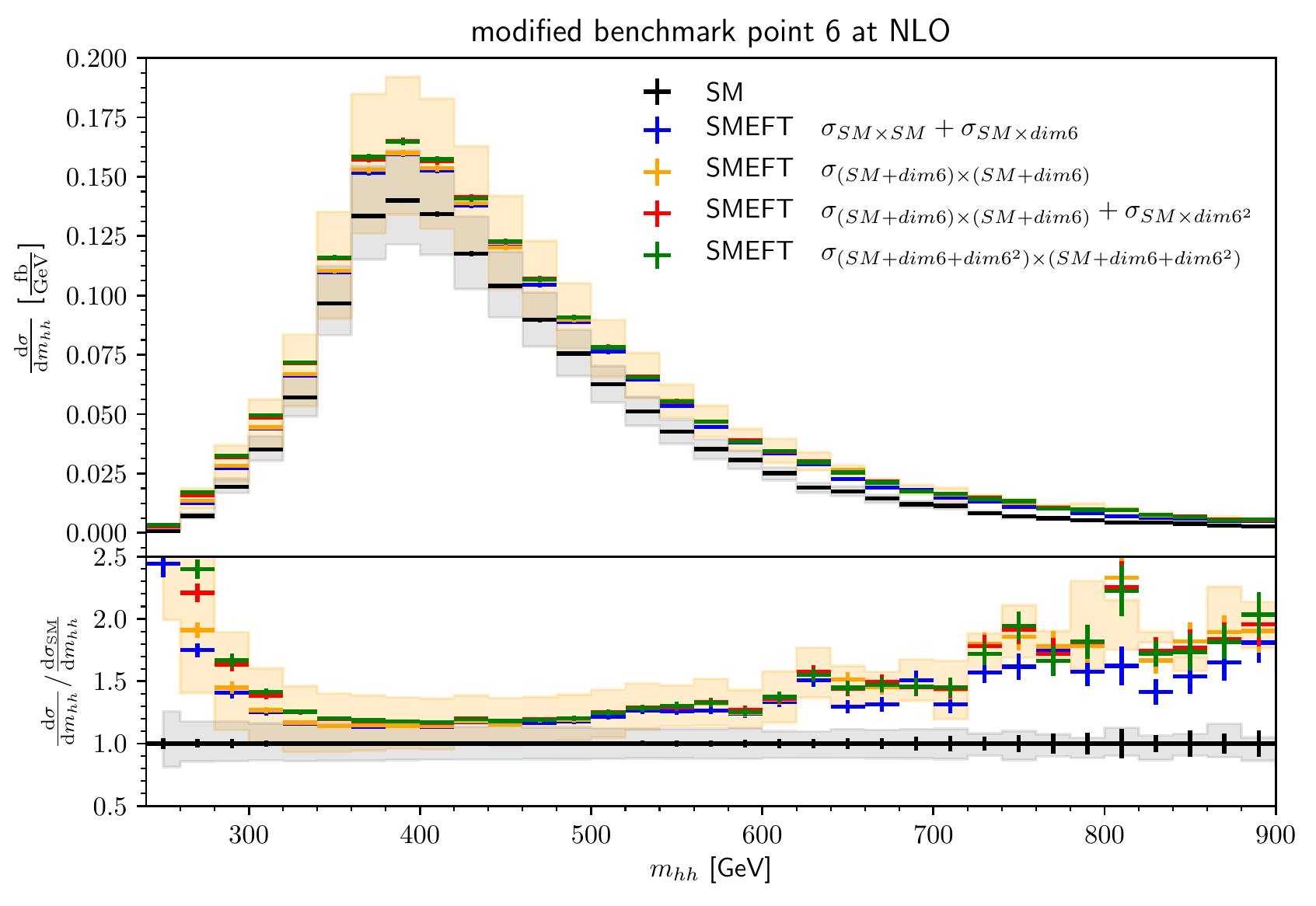}\hspace{2pc}%
\caption{\label{XSdistributions} Differential cross sections for the
  Higgs boson pair invariant mass. Top row: 
$\Lambda=1$\,TeV, bottom row: $\Lambda=2$\,TeV.
Left: benchmark point $1^\ast$, right: benchmark point $6^\ast$ of Table~\ref{tab:benchmarks}. The gray and orange bands show the uncertainty from 3-point scale variations $\mu_R=\mu_F=c\cdot m_{hh}/2$, with $c\in\{\frac{1}{2},1,2\}$, for the SM and SMEFT $\sigma_{(SM+dim6)\times (SM+dim6)}$, respectively.}
\end{figure}


\section{Summary}

We have presented NLO QCD corrections to Higgs boson pair production in combination with a Standard Model Effective Field Theory (SMEFT) parametrisation of effects of physics beyond the Standard Model.
The calculation has been implemented into the  {\tt GoSam+POWHEG} Monte Carlo program framework in a way which allows to choose different options for the truncation of the EFT series and to compare to results in (non-linear) Higgs Effective Field Theory (HEFT).
The results show that a naive translation between HEFT and SMEFT has pitfalls and that the various truncation options can lead to large differences in the theory predictions.

\subsection*{Acknowledgements}

We would like to thank Stephen Jones, Matthias Kerner and Ludovic
Scyboz for collaboration in the $ggHH$@NLO project and Gerhard
Buchalla for useful discussions. Special thanks to Ludovic for providing up-to-date benchmark points.
This research was supported by the Deutsche Forschungsgemeinschaft (DFG, German Research Foundation) under grant 396021762 - TRR 257.

\section*{References}

\bibliographystyle{iopart-num}

\bibliography{gghh_SMEFT}

\end{document}